\documentclass[useAMS,usenatbib]{mn2e}
\usepackage[dvips]{color,graphicx}
\newcommand{\bi}{\bfseries\itshape} 

\title[Shallowed DM cusp slope in clump clusters]{Shallowed cusp slope of dark matter in disc galaxy formation through clump clusters}
\author[S. Inoue \& T. R. Saitoh]{Shigeki
Inoue$^{1,2}$\thanks{E-mail:inoue@astr.tohoku.ac.jp} \& Takayuki R. Saitoh $^{3,4}$
\\
$^{1}$Astronomical Institute, Tohoku University, Sendai 980-8578, Japan\\
$^{2}$Mullard Space Science Laboratory, University College London, Holmbury St. Mary, Dorking, Surrey, RH5 6NT\\
$^{3}$Center for Computational Astrophysics, National Astronomical Observatory of Japan, Mitaka, Tokyo 181-8588, Japan\\
$^{4}$Interactive Research Center of Science, Tokyo Institute of
Technology, 2--12--1 Ookayama, Meguro, Tokyo 152--8551, Japan}
\begin{document}

\date{2011 April 22}

\pagerange{\pageref{firstpage}--\pageref{lastpage}} \pubyear{2002}

\maketitle

\label{firstpage}

\begin{abstract}
Cusp-core problem is a controversial problem on galactic dark matter haloes. Cosmological $N$-body simulations has demonstrated that galactic dark matter haloes have a cuspy density profile at the centre. However, baryonic physics may affect the dark matter density profile. For example, it was suggested that adiabatic contraction of baryonic gas makes the dark matter cusp steeper. However, it is still an open question if the gas falls into the galactic centre in smooth adiabatic manner. Recent numerical studies suggested that disc galaxies might experience clumpy phase in their early stage of the disc formation, which could also explain clump clusters and chain galaxies observed in high redshift Universe. In this letter, using numerical simulations with an isolated model, we study how the dark matter halo responds to these clumpy nature of baryon component in disc galaxy formation through the clump cluster phase. Our simulation demonstrates that such clumpy phase leads to a shallower density profile of the dark matter halo in the central region while clumps fall into the centre due to dynamical friction. This mechanism helps to make the central dark matter density profile shallower in the galaxies whose virial mass is as large as $5.0\times10^{11}$ M$_\odot$. This phenomenon is caused by reaction to dynamical friction of the stellar clumps against the dark matter halo. The halo draws the clumps into the galactic centre, while kinematically heated by the clumps. We additionally run a dark matter only simulation excluding baryonic component and confirm that the resultant shallower density profile is not due to numerical artifact in the simulation, such as two-body relaxation.

\end{abstract}


\section{Introduction}
Cosmological $N$-body (dark matter only) simulations with a large number of particles have been carried and demonstrated that the density profiles of the dark matter (DM) halo have cusps at their central region \citep[e.g.][]{nfw:97,fm:97,fm:01,kkb:01,nhp:04,swv:08,nls:10,imp:11}. However, it has been observationally suggested that DM haloes of dwarf and low surface brightness (LSB) galaxies seem to have a central constant density region (cored halo), not a central cusp (\citealt[][and references therein]{gww:07,d:10}; \citealt{obb:10}). 

In the real Universe, the galactic DM haloes might be affected by baryonic physics. Recently, some studies suggested that a probable mechanism to turn the \textit{cusp} into the \textit{core} is strong supernova (SN) feedback following bursty star formation in dwarf galaxies \citep{nef:96,mm:04,rg:05,mwc:08,g:10,pgb:10,obg:10,pg:11}. Others proposed core creation by angular momentum transfer from gas clouds, globular clusters, dwarf galaxies \citep{esh:01,tls:06,gmr:10,cdw:11} and galactic bar \citep{wk:02,es:02}. These baryonic structures lose angular momentum by dynamical friction against DM halo. Then, these studies suggested that the central DM cusp is dynamically heated and smoothed out by reaction of the dynamical friction.

However, most of these discussions of core creation were concentrated on less massive galaxies of which masses are $\hspace{0.3em}\raisebox{0.4ex}{$<$}\hspace{-0.75em}\raisebox{-.7ex}{$\sim$}\hspace{0.3em}10^{9-10}$ M$_\odot$. It is because, unlike massive systems, relatively shallow potential wells of these systems are susceptive to dynamical perturbations \citep{ml:99,d:11}.

Since the systems of dwarf and LSB galaxies have a high mass-to-light ratio even in the central region, the DM profile is more promisingly measurable by kinematics of baryon tracers. On the other hand, most galaxies are dominated by baryon in their central region. Therefore, it is difficult to clarify DM density profile by any observation. However, although there have been fewer studies, \citet{sma:08} has discussed from observation of rotation velocity curves that high surface brightness disc galaxies whose types range from Sab to Im also favors the cored DM halo rather than the cuspy density profile, as well as dwarf and LSB galaxies. Contrary to the result of \citet{sma:08}, on the other hand, some theoretical studies have proposed that central density profile of DM halo of massive galaxies become \textit{steeper} rather than shallower by adiabatic contraction of baryonic gas \citep[e.g.][]{bff:86,gz:02}. 

The theory of adiabatic contraction assumes slow and smooth gas infall. However, some galaxies observed at high redshift Universe have displayed \textit{clumpy} stellar structures in their system, which are referred to as clump clusters and chain galaxies \citep[e.g.][]{bar:96,eeh:04,eem:09,gnj:11}. Numerical simulations suggested that since the highly gas-rich disc is expected in early stage of the disc galaxy formation, the clumpy structures form due to the instability of accreting gas \citep[e.g.][]{n:98,n:99,isg:04,isw:04,bee:07,atm:09,cdb:10,cdm:11,abj:10}. They suggested that these clumpy galaxies would evolved into current disc galaxies. In this scenario, the stellar clump has much larger mass ($\sim10^{5-8}$ M$_\odot$) than ordinary star clusters \citep[e.g.][]{eem:09}. Therefore, the clumps are expected to fall into the galactic centre by dynamical friction. This clumpy manner of baryon accretion may be far from adiabatic. Therefore, it is an open question how the DM responses to such clumpy disc formation.

Although some studies mentioned above have investigated the core creation due to angular momentum transfer in various cases of dynamical heating sources, it is yet to be studied how the DM halo reacts to the disc galaxy formation through clump cluster phase. In this study, we run a numerical simulation for disc galaxy formation in an isolated DM halo and study how DM halo cusp would be affected in disc formation through the clump cluster phase. 

\section{Simulation}
We use an $N$-body/Smoothed Particle Hydrodynamics (SPH) code, {\tt ASURA} \citep{sdk:08,sdk:09}. Gravitational interaction is calculated by parallel Tree with GRAPE (GRAvity PipE) method \citep{m:04}, where we use Phantom-GRAPE (Nitadori et al. in prep.). Gas dynamics is handled by the standard SPH \citep{m:92,s:10}. The number of neighbouring particles of each SPH particle is set to 32$\pm$2. Equations of motion of particles are integrated by the leap-frog method with the individual time-step scheme. We here adopt the time-step limiter \citep{sm:09} which enforces the neighboring particles to have similar time-steps so that SPH with individual time-steps can solve the evolution of the gas in the shocked regions correctly. In addition, the FAST method \citep{sm:10} is also adopted in the time-integration; this algorithm accelerates the simulations of self-gravitating gas by integrating Hamiltonian split into the gravitational potential and others (kinetic and internal energies) with different time-steps.

 Radiative cooling of gas is solved by assuming an optically thin cooling function depending on metallicity of gas particles, which covers a wide temperature range of $10-10^8$ K \citep{wps:09}. Radiative heating due to far-ultraviolet radiation and type-II SN feedback are also taken into account. Unlike feedback implementations of e.g. \citet{tc:01,ssk:06}, we do not halt the radiative cooling in thermalized regions due to SNe. In modeling star formation and the SN feedback, we adopt a simple stellar population approximation with the Salpeter initial mass function (IMF) \citep{s:55} in which stellar mass ranges from $0.1$ M$_\odot$ to $100$ M$_\odot$. A gas particle spawns a stellar particle whose mass is set to have 1/3 of the original gas particle under the criteria: 1) $\rho_{gas}>$100 $atm$/cm$^3$, 2) temperature $T<100$ K, 3) $\nabla\cdot${\bi v}$<0$ and 4) there is no SN around the particle. The local star formation rate is assumed to follow the Schmidt law \citep{s:59}, with the dimensionless star formation efficiency $C_* =0.033$. Since in our simulation we assume that stars form from quite dense and cold gas, the global properties of star formation such as star formation rate or the Schmidt-Kennicutt relation \citep{s:59,k:89} is fairly insensitive to the adopted value of $C_*$ \citep{sdk:08,sdk:09}. We assume that stars heavier than 8 M$_\odot$ cause type-II SNe and thermally heat up ambient gas.

Our initial condition follows the spherical model of \citet{kmw:06,kmw:07} that was used to study the formation of disc galaxies in an isolated environment. We assume an equilibrium system with Navarro-Frenk-White profile \citep[][hereafter NFW]{nfw:97} with a virial mass $M_{\rm vir} = 5.0\times10^{11}$ M$_\odot$, a virial radius $r_{\rm vir} = 1.67\times10^2$ kpc and a concentration parameter, $c=6.0$. Baryon mass fraction of the system is set to 6 per cent. At the initial phase, all of baryon is in the gas phase. Density profiles of DM and gas components are initially the same. The gas is initially primordial (zero-metal gas) and assumed to have virial temperature. The gas sphere has a rotation following a specific angular momentum distribution of $j\propto r$ \citep{bdk:01} normalized by a spin parameter of $\lambda=0.1$, while the DM halo does not rotate. The DM halo is represented by $10^7$ collisionless particles with a softening length of $\epsilon_{DM} = 8$ pc, whereas the gas in the halo is expressed by $5.0\times10^6$ SPH particles with $\epsilon_{gas} = 2$ pc. Star particles formed from gas particles also have the same softening length as the gas particle. 

In addition to the fiducial run, we perform a pure $N$-body run in order to make sure that change in the central density profile is not due to numerical artifact, such as two-body relaxation. The $N$-body run uses the same density profile as that of the fiducial run, which is represented by only the DM particles with the same resolution with no net rotation.

\section{Results}
\begin{figure*}
  \begin{minipage}{170mm}
   \includegraphics[width=170mm]{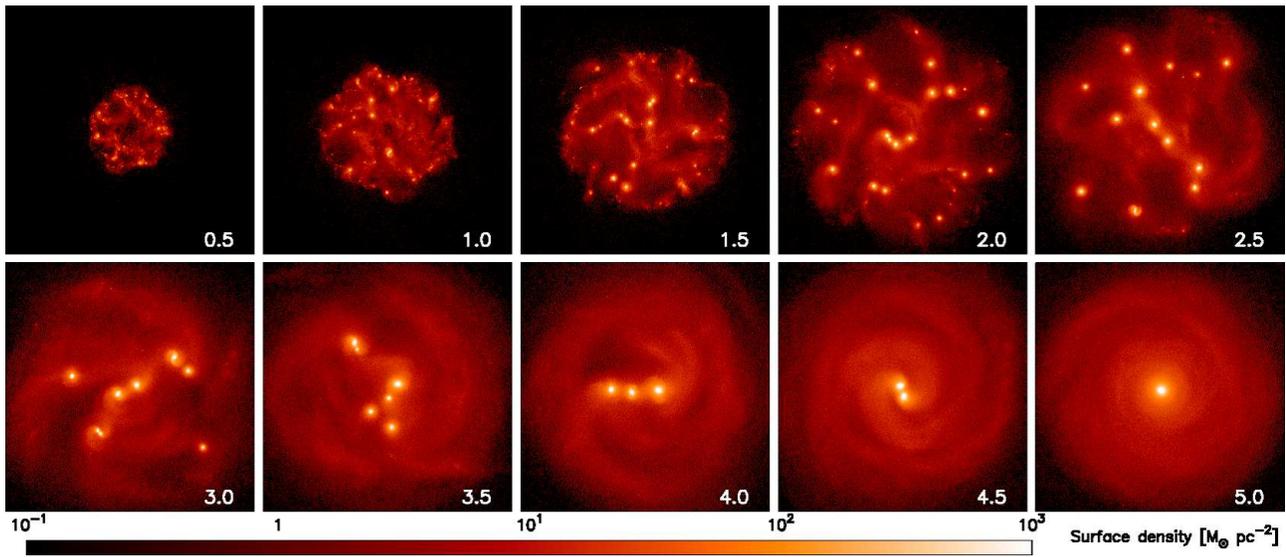}
   \caption{Stellar surface density maps from face-on view in central $60\times60$ kpc region. Time in unit of Gyr is indicated on right bottom in each panel.}
   \label{snap}
  \end{minipage}
\end{figure*}

We run the simulation for $6$ Gyr and show snapshots during the simulation in Fig.\ref{snap}. The galaxy simulated settles into a stable state after $5$ Gyr, in which the density map unchanges.

At $t\hspace{0.3em}\raisebox{0.4ex}{$<$}\hspace{-0.75em}\raisebox{-.7ex}{$\sim$}\hspace{0.3em}2$ Gyr, many clumps form in the disc while the surrounding gas is falling onto the disc. In this phase, since the disc is highly gas-rich, radiative cooling enhanced in the dense gas induces self-gravity instability and results in the clump formation. This clumpy phases in the initial gas rich disc are seen in previous numerical studies and suggested to explain the clump clusters observed at high redshift Universe \citep[e.g.][]{n:98,n:99,isg:04,isw:04,bee:07,atm:09,cdb:10,cdm:11,abj:10}. In our simulation model, the gas infall ceases around $t\sim2$ Gyr. The clumps decrease in number because of merger with one another and tidal disruption \citep{sn:93}. Finally, all clumps merge into a single \textit{``clump-origin''} bulge as first suggested by \citet{n:98,n:99}. The galactic disc in our simulation is formed from clumps disrupted by galactic tide, while clumps migrate to the galactic center. At $t=6$ Gyr, eighty per cent in gaseous mass had turned into stars. There is little gas in halo region.

The bulge has a mass of $4.5\times10^9$ M$_\odot$, a half-mass radius of $r_e=450$ pc and a mean surface brightness within $r_e$ of $I_e=3.5\times10^3$ L$_\odot$/pc$^2$ under an assumption that stellar mass-to-light ratio is 1.0. These values are consistent with an $r_e$-$I_e$ relation observed among bulges and elliptical galaxies \citep[e.g.][]{bbf:97}. We find the transition radius between the disc and bulge components at $R\sim1$ kpc on the surface density. \citet{bem:09} has suggested that the disc formed from the clumps is a thick disc. In our simulation, a thin disc hardly forms and a thick disc dominates over the thin disc. The thick disc has a scale height of 1.41 kpc, a scale radius of 9.85 kpc and a nearly exponential density profile. We will report all the details of the bulge and disc formation in our simulation in forthcoming paper (Inoue \& Saitoh, in prep).

\begin{figure}
  \includegraphics[width=90mm]{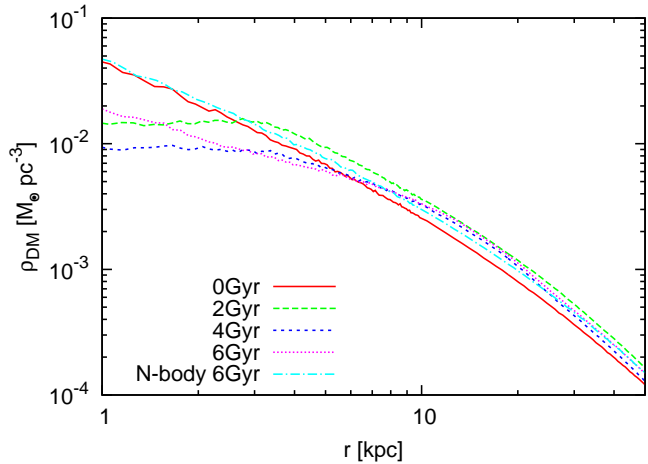}
 \caption{Time-evolution of density profiles of the DM halo. The line labelled as 'N-body 6Gyr' indicates result of the DM only simulation. The centre of the halo is defined to be the centre of mass of the DM particles.}
 \label{profs}
\end{figure}

In the disc formation scenario through clump cluster, large clumps are sucked into galactic centre by dynamical friction against baryon and DM particles. Therefore, the DM halo must be kinematically heated by the clumps as reaction to the dynamical friction. We show time-evolution of DM density profile in this simulation in Fig.\ref{profs}. As seen in the figure, DM density profile becomes nearly constant at the central region, $r\hspace{0.3em}\raisebox{0.4ex}{$<$}\hspace{-0.75em}\raisebox{-.7ex}{$\sim$}\hspace{0.3em}3~{\rm kpc}$, i.e. cored. The cored region expands until $t=4$ Gyr. However, the central cusp \textit{revives} at $t=6$ Gyr due to the bulge formation: settlement of a heavy bulge makes potential well deeper and drags matter into the centre. The resultant cusp slope of DM density profile is significantly shallower than the initial state and central density concentration is greatly reduced \citep{ebe:08}.
 
We perform a pure $N$-body run additionally in order to discuss artificial effect such as two-body relaxation. This $N$-body system must be in the equilibrium state as long as the relaxation process does not work. The result is superposed on Fig.\ref{profs}. The density profile of the $N$-body run is consistent with the initial state and unchanged even after $6$ Gyr, demonstrating that the shallowed DM density profile in the fiducial run is not caused by the artificial relaxation.

\section{discussion and summary}
In this letter, we first discussed the cusp-core problem in the formation stage of a disc galaxy through clump cluster. As seen in Fig.\ref{profs}, the central DM density slope is clearly shallowed by the stellar clumps falling into the galaxy centre. This would be caused by reaction to dynamical friction on the clumps. This mechanism is basically the same as in the cases of other dynamical sources discussed in the previous works \citep{wk:02,es:02,esh:01,tls:06,gmr:10,cdw:11}. The result suggests that DM haloes would be cored in the clump clusters and chain galaxies in the high redshift Universe. In our simulation, the DM cusp has been smoothed out in the early stage of their formation (the clump cluster phase) and a shallowed DM cusp remains at the later epoch. This process can be effective also in current disc galaxies like the Milky Way, if it used to be a clump cluster and if there is no extra dynamical perturbation.

Our simulation model is missing some physical processes possible in cosmological context, such as galaxy merger and accretion flow. However, at first, recent cosmological simulations has demonstrated that disc galaxies in high-redshift Universe are mainly fed by smooth accretion flow, not by galactic merger \citep[e.g.][]{d:09}. Then, we expect that the lack of the merger events would not spoil reasonableness of our simulation. Next, these cosmological simulations showed that the smooth accretion is narrow cold gas stream, which is different from our simulation model: isolated collapse model. Nevertheless, simulations using the isolated collapse model can also successfully reproduce the clumpy nature of high-redshift galaxies \citep{n:98,n:99,isg:04,isw:04,abj:10}. Therefore, we expect that the isolated collapse model would not be a seriously incorrect assumption for qualitative discussion.

Our simulation result indicates a possibility that progenitors of current disc galaxies have a cored DM halo even before the disc forms. However, our simulation model is lacking the later accretion to form a thin disc \citep{bem:09}. Such a late accretion may also affect the DM density. For example, the dynamically cold thin disc may forms spiral arms and/or galactic bar, which can form a pseudobulge \citep[e.g.][]{kk:04}. It is also possible that the pseudobulge steepens the DM density slope. Meanwhile, the galactic bar may also flatten the density slope \citep{wk:02,es:02}. Then, we have to note that it is still a open question whether DM haloes of current spirals are cored or cuspy.

The cored DM density profile brings important influences on dynamics; rotation curve, orbit of star clusters \citep{tw:84,w:89,hg:98,gmr:06,rgm:06,i:09,i:11}, tidal effect on the clusters \citep{hdf:03}, moreover, intensity of $\gamma$-ray emission (if annihilating radiation of DM was possible and observable). As mentioned in \S 1, \citet{sma:08} has discussed that high surface brightness disc galaxies favour the cored DM halo rather than the NFW cuspy profile. The $\gamma$-ray emission from the Milky Way centre has not been clearly observed yet by early result of \textit{Fermi} telescope (\citealt{j:09,vm:09}, although see \citealt{gh:09}) although this only means that positive evidence of enough high density of DM has not been observed in the Galactic Centre, since there are some other physics to explain the no detection.

In our simulation, the bulge formation reconstructs the DM cusp at the final state although the slope is shallower than the NFW cusp. This result implies that DM density profiles in current disc galaxies would depend on size (mass) of the bulge: even if clump clusters make the central DM density cored, formation of a bigger bulge leads to the revival of steeper central cusp. Then, if we naively assume that all bulges are formed by clump mergers in a clump cluster, early-type spirals may have a cuspy DM profile since they tend to have a large bulge. On the other hand, since late-type spirals, like LSB galaxies, tend to have a smaller bulge (or be bulgeless), they may have a shallower cusp (or cored) DM halo. However, clump clusters may inevitably lead to formation of the clump-origin bulge. Therefore, the galaxy formation scenario through clump cluster phase may not be eligible for formation of the late-type and bulgeless galaxies. However, \citet{gnj:11} has recently observed very strong outflow from the stellar clumps in clump clusters. \citet{gng:10} adopted such strong outflow model in their simulation and suggested that the clumps would be destroyed due to strong outflow from the clumps before reaching the galactic centre. In such a case, some disc galaxies forming through clump cluster phase may be able to evade the bulge formation and the central cores may remain. 

However, the clumps in our simulation seem long-lived, in contrast with the short-lived clumps suggested by e.g. \citet[][]{gng:10,gnj:11}. The SN feedback model adopted in our simulation can not reproduce such destructive outflow. Although to investigate what is the cause of the difference is beyond the scope of this letter, we speculate that IMF may be largely different from the Salpeter IMF in gas-rich circumstance of the clumpy galaxies. If the clumpy galaxies have a top-heavy IMF, the observational result of such powerful outflow may be explainable. Addtionally, it is also discussed that SN feedback in simulations can not reach very high temperature under the simple stellar population approximation \citep[e.g.][]{sm:10}.

\section*{Acknowledgments}

We thank the anonymous reviewer for important suggestions that helped improve the letter. We are grateful to Daisuke Kawata for polishing up the paper. S.I. is financially supported by Research Fellowships of the Japan Society for the Promotion of Science (JSPS) for Young Scientists. The numerical simulations reported here were carried out on Cray XT-4 kindly made available by CfCA (Center for Computational Astrophysics) at National Astronomical Observatory of Japan. This project is supported by HPCI Strategic Program Field 5 `The origin of matter and the universe'.


\end{document}